\documentstyle[psfig]{mn}
\title[Opening angle in AGN]{The luminosity dependence of opening
angle in unified models of active galaxies.}
\author[C.M.Rudge \& D.J.Raine]{C.M.~Rudge$$ and D.J.~Raine$$
\\$$Astronomy Group, University of Leicester, University Road,
Leicester, LE1 7RH, UK.}
\begin{document}
\maketitle
\begin{abstract}
In unified models of active galaxies the direct line of sight to the
nucleus is unobscured only within a certain cone of directions. An
opening angle for this cone is usually estimated by methods such as
the overall ratio of Seyfert 1s to Seyfert2s, the latter assumed to be
obscured versions of the former. Here we shall show, as has often been
suspected, that the opening angle of the cone depends on the luminosity
of the central source, with higher luminosities corresponding to larger
opening angles. This conclusion depends only on the assumption that
the width of the broad emission lines at a given luminosity is a
measure of inclination angle, an assumption that is supported by
observation in radio-loud systems. On the other hand we show that the
scatter in X-ray spectral index is not primarily an effect of viewing
angle, in contrast to what might be expected if the scatter on the
spectral index versus luminosity relation were a consequence of absorption
in the obscuring material. The observed correlation between linewidth
and spectral index appears to be a further consequence of the
dependence of opening angle on luminosity.
\end{abstract}

\section{Introduction}

The basic structure of an active galactic nucleus (AGN) includes a
central continuum source and emission line gas, which, for some lines
of sight is blocked from direct view by obscuring material of some
form, possibly having the geometry of a torus. Viewed along the
opening cone of the torus such systems appear as unobscured Seyfert 1
nuclei (in the radio-quiet case), while from greater inclinations to
the axis of the torus the broad lines cease to be directly visible and
they appear as Seyfert 2s. These are analogous to type 1 and 2 QSOs in
the radio-loud case. It has been clear from the outset that this
simple picture, in which the opening angle of the torus is fixed for
all systems, is unlikely to be true \cite{Antonucci93}. In this paper
we shall show that the available data on the linewidth distribution of
the broad emission lines can be interpreted in terms of an opening
angle that increases with luminosity.

The conclusion depends on the assumption that we can use the broad
line widths at given luminosity to measure the angle of inclination of
the axis of the obscuring matter to the line of sight. This implies
both that the broad line region (BLR) is axisymmetric and that its
axis coincides with that of the obscuring material. In the case of
radio-loud systems the axis of radio emission can be shown to
correlate with broad line widths \cite{Wills86} implying that the BLR
is axisymmetric and aligned orthogonally to the radio axis. In
radio-quiet systems we have shown \cite{RR98} that the distribution of
scatter on the linewidth -- luminosity relation can be accounted for
in terms of inclination. We examine these points further in
section~\ref{sec:axisym}.

At first sight we should be able to use our relation between
linewidth, luminosity and inclination angle to measure the inclination
of individual systems. Despite the agreement with the statistical
distribution however, we appear to find problems carrying this out,
specifically that for a number of galaxies there is no solution for
the angle. The reason for this can be readily seen if we bin the data
into luminosity ranges and allow the distribution in $\sin i$ of the
objects to be determined by the data. We find that $i$ is restricted
to a range $i < i_{\ast}$ where $i_{\ast}$ increases with luminosity
$L$ (section~\ref{sec:results}). We take $i_{\ast}$ as a measure of
the opening angle in the unified model. Our original assumption that
systems have random inclination, i.e.\ $\sin i$ is uniformly
distributed in $[0, \pi/2]$, is therefore not valid. We expect that
the distribution of $\sin i$ is constant at each luminosity, but the
limited size of the data set does not show this clearly.

In section~\ref{sec:FWHM_ax} we consider the proposed anti-correlation
between broad line widths and the X-ray spectral index, $\alpha_{\rm
x}$ \cite{BG92,Wandel98,P97}. We show that this is not primarily
driven by a dependence of $\alpha_{\rm X}$ upon viewing angle, as
might be expected from the dependence of FWHM upon orientation. The
observed anti-correlation is at least partially due to the increased
range of viewing angles at higher luminosities.

\section{Evidence for axisymmetry in the Broad Line Region}
\label{sec:axisym}

The radio power in radio galaxies is generally accepted to be an
indicator of viewing angle to the central source, with the flat
spectrum core dominant in face-on systems and the steep spectrum lobes
dominant in edge-on systems. The ratio of core to lobe radio power, $R$,
correlates with the width (FWHM) of H$\beta$ in the sense that the
broadest lines are seen in more edge-on systems \cite{Wills86}. Wills
\& Brotherton \shortcite{Wills95} develop this further with the
introduction of a new parameter, $R_{\nu}$. This is defined to be the
ratio of radio core luminosity at 5GHz rest frequency to the optical
V-band luminosity - improving the measure of orientation. They
show that $R_{\nu}$ has a stronger correlation than $R$ to the jet
angle in a sample of 33 FR II sources. Further they show that using
$R_{\nu}$ rather than $R$ also improves the correlation with
FWHM$_{\rm H\beta}$ for both the Wills \& Browne \shortcite{Wills86}
objects and a new sample of low-$z$ quasars \cite{Brotherton96}, thus
strengthening the case for an axisymmetric BLR in radio-loud
systems. This evidence is supported by the correlation between FWHM
and $\alpha_{\rm ox}$. The optical continuum is boosted by the jet in
the face-on systems giving a viewing angle dependence for the optical
to X-ray spectrum slope $\alpha_{\rm ox}$.

The case is much less clear for the radio-quiet systems. Here we have
no such obvious inclination indicators as the jet angle. However,
there is little, if any, strong evidence to suggest that the BLR in
radio-quiet systems should be significantly different to that in the
radio-louds. Studies of the distribution of line widths for
radio-quiets and louds show only a small difference in the
distributions with the radio-louds having generally wider lines
\cite{Corbin97}. However, the radio-loud systems in this sample have a
higher average luminosity. Since higher luminosity systems have on
average broader lines, at least for C{\sc iv} and H$\beta$, the result
is what we would expect if the systems are drawn from a common
population. Marziani et al.\ \cite{Marziani96} consider in more detail
the differences between the profiles of H$\beta$ and C{\sc iv} lines
in radio-loud and radio-quiet systems. They conclude that the line
profile properties indicate that the BLR in radio-louds is not the
same as that in radio-quiets. We shall discuss their findings in the
context of the results of this paper in section
\ref{sec:discussion}. Boroson \shortcite{Boroson92} argues against a
viewing angle dependent picture of radio-quiet AGN in which
\emph{both} the continuum and line emission are axisymmetric by
consideration of the lack of correlation between the equivalent width
of [O{\sc iii}] and the FWHM of H$\beta$. The sample is selected by UV
excess which may produce a bias against edge-on objects. In addition,
according to the picture to be developed here, the range of angles for
a low luminosity radio-quiet sample may be rather small, so the
evidence may not be conclusive. In any case, all we require here is
that the broad line region kinematics be axisymmetric, not that the
illuminating continuum should be too.

From a theoretical point of view spherical BLRs dominate the
literature. However, to provide the observed variations in profile
shape with width \cite{Stirpe91} such systems have to be quite
complex. For example Robinson \shortcite{Robinson95} uses a changing
radial depth and radial power laws for the velocity and emissivity of
the gas to obtain the range of profile shapes. Nevertheless, while
this provides an adequate account of individual systems, it is not
clear whether models of this type can fit the linewidth
distribution. A number of simple flow geometries in spherical systems
are ruled out by detailed observation. For example, for NGC 3516, Goad
et al.\ \shortcite{Goad99} exclude both radial flows at constant
velocity and Keplerian motion. Spherically symmetric systems also do
not appear to be able to account for the change in profile shape with
line width or the range of widths at each luminosity needed to be able
to account for the distribution of linewidths \cite{RR98}. Simple disc
geometries are also ruled out by consideration of the change in
profile shape with linewidth \cite{Stirpe91}. However more complex
systems such as the dual winds model of Cassidy \& Raine
\shortcite{CR96} or the VBLR--ILR model of Wills et al.\
\shortcite{W93} adopted by Puchnarewicz et al.\ \shortcite{P97} are
able to predict the change in profile shape with linewidth required by
observation \cite{Stirpe91}.

An alternative approach \cite{Gaskell82,Dumont90} envisages a two-zone
model which distinguished between high and low ionization
lines. Originally prompted by the systematic blueshift of C{\sc iv}
relative to the Balmer lines, which suggested origins in regions of
different kinematics, and by considerations of energy balance, the
idea has received some support from reverberation mapping. The
H$\alpha$ transfer function peaks away from zero delay (for example in
NGC 3516, Wanders\& Horne 1994\nocite{Wanders94}), consistent with a
flattened cloud distribution, while the C{\sc iv} response is
immediate (for example NGC 5548, Korista et al.\
1995\nocite{Korista95}), implying material in the line of sight. As a
result the high ionization region (HIL) is often taken to be
spherical, although alternative geometries are also consistent with
echo mapping \cite{Marziani96}. Although elsewhere we have argued that
the statistical properties of the H$\beta$, Mg{\sc ii} and C{\sc iv}
linewidth distributions do not indicate substantial differences
between HIL and LIL geometries, the arguments in this paper depend
only on the validity of some orientation effect in both H$\beta$ and
Mg{\sc ii}. While Mg{\sc ii} emission is often taken to be associated
with the Balmer lines, and certainly arises from a region more
extended than that producing C{\sc iv} (from observed
cross-correlation functions), it should perhaps be born in mind that
we lack independent evidence for the geometry of the Mg{\sc ii}
emitting region.

The current work therefore provides a further self-consistency
argument for axisymmetry in the BLR.

\section{Line width distribution}
\label{sec:LWD}

One of the reasons for developing a viewing angle dependent model is
the need for some parameter, other than luminosity, upon which the
FWHM of the broad emission lines depends. It is of course possible
that the dominant parameter could be something other than viewing
angle. Perhaps the most obvious choice would be black hole mass,
$M$. However the success of a model \cite{RR98} in which the unknown
parameter varies as a sine function suggests that this is not the
case: it is surely unrealistic to suggest that $M$ has only part of a
sine distribution. Furthermore, we will show in this paper that the
range of values taken by this parameter increases with
luminosity. Assuming that $L\propto M\dot{M}$ then we would expect $M$
to take a smaller, rather than larger, range of values at higher
luminosities.

Thus we assume that, in general, the FWHM, $v$, of a given broad
emission line in an axisymmetric system can be given as a function of
the ionising luminosity and the inclination of the system. This
function can be expanded in spherical harmonics with luminosity
dependent coefficients. In a previous paper \cite{RR98} we showed that
the distribution of linewidths could be reproduced if this function
were taken to be axisymmetric and only the first two terms of this
series were retained, with the coefficients taken to have a common
dependence on luminosity. The FWHM of a given emission line is then
given by
\begin{equation}
	v=(a+b\sin i)L^{\alpha}
	\label{eqn:vil}
\end{equation} 
with the constants $a$, $b$ and $\alpha$ being chosen for each
emission line. The inclination angle, $i$, is the angle of the line of
sight to the axis of the BLR i.e.\ $i=0$ for face-on systems. In Rudge
\& Raine \shortcite{RR98} we took $L$ to be the B-band
luminosity.

Since it is difficult to determine the line of sight angle for
individual systems, at least with any accuracy, we are led to
consider the linewidth distribution rather than the linewidths of
individual objects. Assuming that the inclination of AGN is random
across the sky, then the number of systems at each $v$ is given by
\begin{equation}
	N(v)=\int \frac{\sin i}{\left| \frac{{\rm d}v}{{\rm d}i}
	\right|} \Phi(L) {\rm d}L
	\label{eqn:lwd}
\end{equation}
where $\Phi(L)$ is the luminosity function giving the distribution of
luminosities. In Rudge \& Raine \shortcite{RR98} we used the
luminosity function of Boyle, Shanks \& Peterson
\shortcite{Boyle88}. In the later work on cosmology \cite{RR99} we
used the X-ray luminosity function of Boyle et al.\
\shortcite{Boyle94} and also the optical luminosity function of Pei
\shortcite{Pei95}. In this paper, for convenience, we will again use the X-ray
luminosities. Boyle et al.\ \shortcite{Boyle93} show that $L_{\rm x}
\propto L_{\rm opt}^{0.88\pm0.08}$ and thus using X-ray rather than
optical luminosities will only result in the requirement of a
different value of $\alpha$ in (\ref{eqn:vil}).

Having developed this model and shown its success in accounting for the
linewidth distribution \cite{RR98} we now use it to consider the
viewing angle of individual systems. For this purpose we again use
data from the RIXOS sample \cite{P96,P97}. This provides comprehensive
data on X-ray and optical continuum luminosities as well as spectral
indices, line strengths, equivalent widths and FWHM.
Rearranging (\ref{eqn:vil}) gives
\begin{equation}
	\sin i = \frac{1}{b} \left( \frac{v}{L_{44}^{\alpha}} - a
	\right).
	\label{eqn:sini}
\end{equation}
where $L_{44}$ is the ROSAT 0.5--2\,KeV luminosity in units of
$10^{44}\,{\rm erg\, s^{-1}}$.

The method for finding the orientation for each system relies on
finding a good fit to the linewidth distribution for a sample of
objects and then using the values of $a$, $b$ and $\alpha$ in
(\ref{eqn:sini}). It is therefore important that the distribution is
accurately modelled. We found that it is no longer sufficient to
assume that the given sample has a luminosity distribution which
matches the global luminosity function, or that the distribution of
$\sin i$ is uniform - i.e.\ systems are at random orientation. In such
a situation we found that, although the linewidth distribution could
be matched, it was not possible to generate simultaneously values of
$\sin i$ all lying in the range 0 to 1.

We therefore have to adopt an iterative procedure. In place of
$\Phi(L)$ in (\ref{eqn:lwd}) we use the actual number of systems in
each luminosity bin, $S(L)$, for the selected sample. Similarly we
need to replace $\sin i/ |{\rm d}v/{\rm d}i|$ with $T(\sin i)/ |{\rm
d}v /{\rm d}\sin i|$ where $T(\sin i)$, the number distribution of
$\sin i$, is calculated by consideration of the whole sample in
(\ref{eqn:sini}). We shall find that the $\sin i$ distribution is also
luminosity dependent and so we will in fact use $T(\sin i, L)$, which,
in practice, is determined for a set of discrete luminosity ranges. Clearly
we have to iterate to find $T(\sin i, L)$ and $N(v)$,  by choice of
$a,\ b$ and $\alpha$ in (\ref{eqn:vil}), simultaneously in
the revised linewidth distribution
\begin{equation}
	N(v)= \int \left| \frac{{\rm d}\sin i}{{\rm d}v} \right |
	 T(\sin i ,L) S(L) {\rm d}L.
\label{eqn:lwd2}
\end{equation}

\section{Data from the RIXOS sample}
\label{sec:data}

\begin{figure}
	\psfig{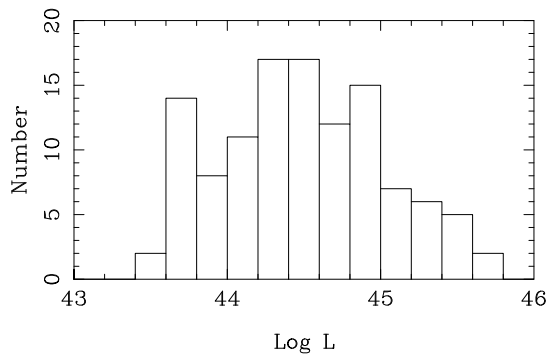}
	\caption{Luminosity distribution of the objects in the
	sample with measured FWHM$_{\rm Mg{\sc ii}}$. $L$ is the total ROSAT
	0.5--2keV band luminosity.}
	\label{fig:Ldist}
\end{figure}
The ROSAT International X-ray/Optical Survey (RIXOS) contains 160 AGN
compiled from serendipitous sources detected in pointed observations
made with \emph{ROSAT}. The optical data was obtained using the Isaac
Newton (INT) and William Herschel Telescopes (WHT) at La Palma. As
well as continuum luminosities and spectral slopes, the data contains
EW, and FWHM for several optical emission lines. Because of the range in
redshift of the objects (0.03--2.92 with most objects at $z<1.0$) it
is clearly not possible to measure these quantities for all the
emission lines with only the WHT and INT. Thus when considering the
sample of AGN in one particular line the sample size of 160 is greatly
reduced and in some cases becomes too small to be of any real use. We
will therefore concentrate on Mg{\sc ii} with a sample size of 113.
Fig.\ \ref{fig:Ldist} shows the luminosity distribution for this
sample.

We note that the X-ray luminosities given in Puchnarewicz et al.\
\shortcite{P97} are not corrected for absorption intrinsic to the
source AGN. However from figure 17 of Puchnarewicz et al.\
\shortcite{P96} we see that 62$\%$ of systems have an absorbing
column, $N_{\rm H}$, of less than $10^{21}\,{\rm cm}^{-2}$ rising to
85$\%$ at $N_{\rm H}<1.5\times 10^{21}\,{\rm cm}^{-2}$. At $N_{\rm
H}=1.5\times 10^{21}\,{\rm cm}^{-2}$ for a standard power law spectrum 
we find that the source luminosity should be $\sim 30\%$ higher than
observed. For the values of $a, b$ and $\alpha$ used here this gives
values of $i$ that are $\sim 10\%$ higher. In most sources the effect
will be much less than this and is therefore neglected.

\section{Results}
\label{sec:results}
\begin{figure}
	\psfig{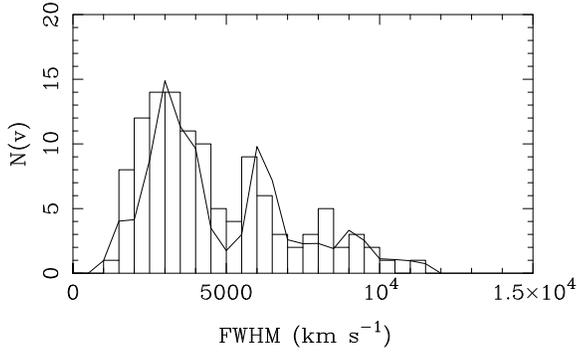}
	\caption{Linewidth distribution curves and data histograms
	for Mg{\sc ii}.}
	\label{fig:lwdfit}
\end{figure}

\begin{figure}
	\psfig{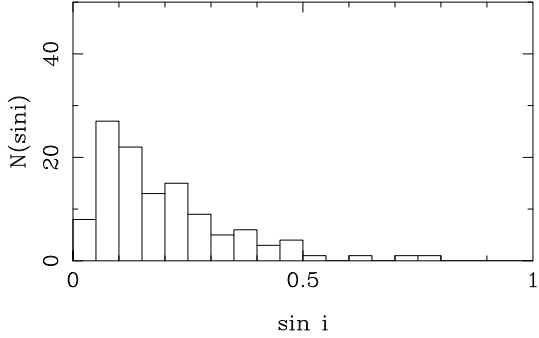}
	\caption{$\sin i$ distribution for Mg{\sc ii} calculated
	using the values of $a$, $b$ and $\alpha$ as in fig.\
	\ref{fig:lwdfit}.}
	\label{fig:sinidist}
\end{figure}
Using (\ref{eqn:lwd2}) and iterating to a solution for $T(\sin i, L)$
we can predict the linewidth distribution for Mg{\sc ii}. We obtain the
values $a=1000 {\rm km\,s^{-1}},\ b=25000{\rm km\,s^{-1}\ and\ }
\alpha=-0.2$. This value for $\alpha$ is well constrained by
consideration of both the Baldwin effect for Mg{\sc ii} and the
observed correlation between FWHM and line equivalent width
\cite{RR98}. The parameters $a$ and $b$ are then constrained tightly
as $b$ gives the spread of the distribution and $a$ its centroid
position. Further constraints are placed on $a$ and $b$ by the
obvious requirement that $0 < \sin i < 1$. Fig.\ \ref{fig:lwdfit}
shows a histogram of the data overlaid with the predicted distribution
curve.

Fig.\ \ref{fig:sinidist} shows the self-consistent distribution of
$\sin i$ calculated from (\ref{eqn:sini}). Since there are no strong
angle-dependent selection effects, this distribution is at first sight 
incompatible with our interpretation of $i$.
\begin{figure*}
	\mbox{\psfig{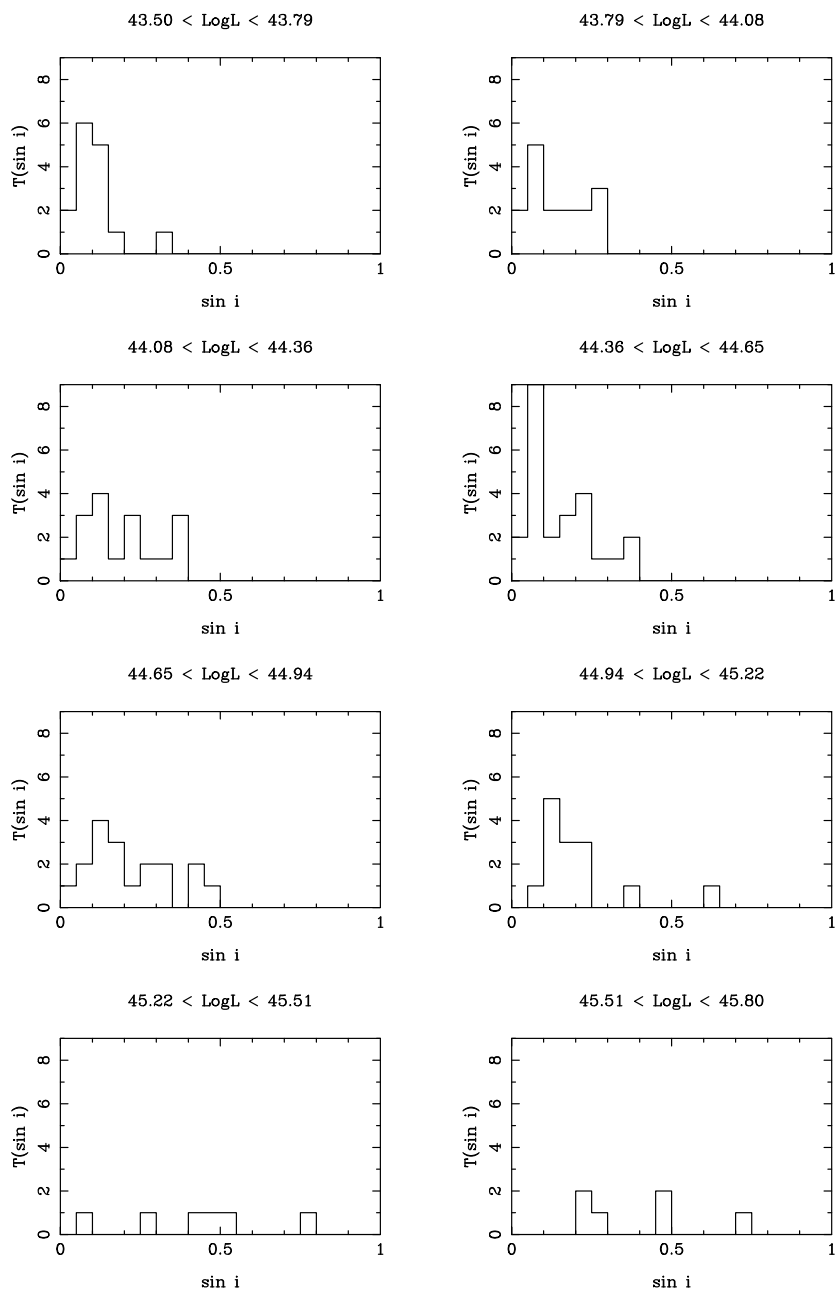}}
	\caption{Change in distribution of $\sin i$ with
	luminosity for Mg{\sc ii}.}
	\label{fig:Lbinsini}
\end{figure*}
However, fig.\ \ref{fig:Lbinsini} shows how the distribution of
$\sin i$ changes with luminosity, the division being into eight
equally sized bins on a logarithmic scale between $\log L = 43.5$ and
$\log L = 45.8$.
\begin{figure}
	\psfig{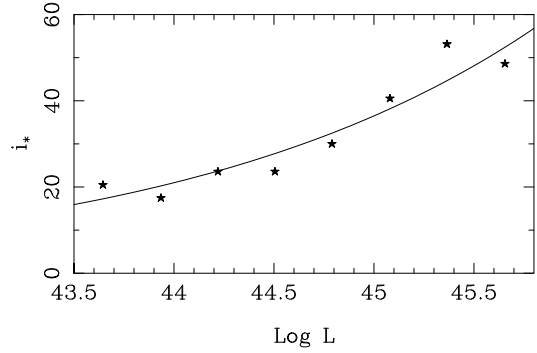}
	\caption{Relation between $i_{\ast}$ and $\log L$ found by
	taking average luminosity and calculated $\sin i_{\ast}$ in
	each luminosity bin. The curve is $i_{\ast} = 21.0 L_{44}^{0.24}$.}
	\label{fig:istarL}
\end{figure}

It is clear that the maximum value of $\sin i$, and therefore
$i_{\ast}$ increases with luminosity. Within each bin the $\sin i$
distribution is more uniform over the range $i<i_{\ast}$ than for the
sample as a whole. From the dependence of $i_{\ast}$ upon $L$ shown in fig.\
\ref{fig:Lbinsini} we fit a relation of the form
\begin{equation}
	i_{\ast} = 21.0 L_{44}^{0.24}\ {\rm degrees}
\label{eqn:istarL}
\end{equation}
Finally table \ref{tab:rixosmgii} shows the sources used,
their luminosities, linewidths and the calculated inclination
angle.

\begin{table}
\caption{Sample data for the RIXOS objects with measured FWHM$_{\rm
Mg{\sc ii}}$. Note: Field ID (1), Source number (2) and linewidths (5)
are taken from Puchnarewicz et al.\ 1997; $\alpha_{\rm x}$ (3) is from
Puchnarewicz et al.\ 1996; $\log L$ for the ROSAT 0.5--2\,keV band is
from E.M. Puchnarewicz (private communication) and $\sin i$ (6) is as
calculated by the method outlined in this paper.}
\[
\begin{array}{cccccc}
{\rm FID} & {\rm Snum} & \alpha_{\rm x} & \log L &
{\rm FWHM_{Mg{\sc ii}}} & \sin i \\
(1) & (2) & (3) & (4) & (5) & (6) \vspace{5pt} \\
110 &   1   &   1.687  &   43.816  &   1945.5 &    0.031 \\
110 &   8   &   1.005  &   44.473  &   7974.5 &    0.356 \\
110 &  34   &   0.033  &   44.363  &   4366.6 &    0.166 \\
110 &  50   &   0.989  &   44.779  &   4088.5 &    0.194 \\
122 &  14   &   1.621  &   43.893  &   2511.7 &    0.055 \\
123 &  41   &   0.763  &   45.159  &   2606.3 &    0.137 \\
123 &  42   &   1.045  &   43.664  &   3035.8 &    0.064 \\
123 &  46   &   0.047  &   44.464  &   5654.4 &    0.240 \\
123 &  66   &   1.030  &   43.943  &   8022.6 &    0.272 \\
123 &  85   &   1.885  &   44.644  &   1651.0 &    0.048 \\
125 &  14   &   0.956  &   45.540  &   4164.4 &    0.298 \\
125 &  17   &   1.477  &   44.271  &   2486.1 &    0.072 \\
133 &  22   &   0.802  &   45.398  &   4293.5 &    0.286 \\
205 &  11   &   1.430  &   44.411  &   2398.3 &    0.075 \\
205 &  12   &   1.430  &   45.185  &   3463.3 &    0.199 \\
205 &  22   &   1.645  &   44.237  &   3958.3 &    0.136 \\
205 &  34   &   1.201  &   44.198  &   5729.7 &    0.211 \\
206 &   6   &   0.672  &   44.279  &   3510.4 &    0.119 \\
206 &   9   &   0.487  &   44.371  &   2700.0 &    0.088 \\
206 & 507   &   0.979  &   43.712  &   3808.7 &    0.093 \\
206 & 522   &   1.373  &   44.338  &   2331.7 &    0.068 \\
208 &   2   &   1.278  &   44.059  &   2690.0 &    0.070 \\
208 &  55   &   0.767  &   45.221  &   9816.8 &    0.649 \\
211 &  30   &   0.951  &   45.182  &   1993.3 &    0.097 \\
211 &  35   &   1.123  &   43.754  &   5584.9 &    0.159 \\
212 &   6   &   0.836  &   44.597  &   3719.0 &    0.155 \\
212 &  16   &   0.954  &   44.345  &   9069.2 &    0.385 \\
212 &  25   &   1.373  &   44.567  &   6667.0 &    0.306 \\
213 &   7   &   0.713  &   44.007  &   3300.6 &    0.092 \\
213 &  11   &   -0.49  &   44.309  &   7363.2 &    0.299 \\
213 &  17   &   1.415  &   43.984  &   4107.5 &    0.123 \\
213 &  19   &   1.415  &   44.117  &   1994.3 &    0.044 \\
213 &  20   &   1.015  &   44.410  &   2501.2 &    0.080 \\
215 &   1   &   0.977  &   45.621  &   2914.0 &    0.205 \\
215 &  19   &   1.375  &   44.396  &   1592.6 &    0.036 \\
215 &  32   &   1.208  &   44.001  &   2696.4 &    0.067 \\
216 &   7   &   0.834  &   44.337  &   5526.4 &    0.218 \\
217 &   3   &   1.214  &   44.590  &   4641.9 &    0.203 \\
217 &  21   &   1.261  &   44.023  &   4852.8 &    0.156 \\
217 &  34   &   1.332  &   44.976  &   3888.3 &    0.203 \\
217 &  35   &   0.921  &   43.640  &   4646.8 &    0.117 \\
217 &  59   &   1.117  &   44.243  &   9304.6 &    0.376 \\
218 &   1   &   0.854  &   44.318  &   5028.3 &    0.192 \\
218 &   9   &   1.508  &   44.222  &   4071.8 &    0.140 \\
218 &  27   &   1.055  &   44.571  &   3273.2 &    0.130 \\
219 &  15   &   1.586  &   44.944  &   2629.2 &    0.122 \\
219 &  45   &   0.546  &   45.096  &   6590.6 &    0.396 \\
219 &  48   &   0.946  &   44.806  &   8955.0 &    0.479 \\
220 &  13   &   0.919  &   43.785  &   3593.6 &    0.090 \\
220 &  18   &   0.784  &   43.685  &   10127. &    0.310 \\
221 &   2   &   0.939  &   44.445  &   2281.8 &    0.072 \\
221 &  35   &   0.948  &   44.845  &   2401.9 &    0.101 \\
224 & 201   &   1.265  &   45.238  &   8099.2 &    0.532 \\
226 &  41   &   1.079  &   45.333  &   6435.1 &    0.435 \\
226 & 114   &   0.756  &   44.605  &   2279.0 &    0.080 \\
227 &  19   &   1.346  &   45.504  &   1744.9 &    0.099 \\
227 &  37   &   1.209  &   45.528  &   6654.0 &    0.497 \\
227 & 513   &   0.912  &   44.761  &   3553.8 &    0.161 \\
\end{array}
\]
\end{table}
\begin{table}
\contcaption{Data for objects in RIXOS sample with measured FWHM$_{\rm 
Mg{\sc ii}}$.}
\[
\begin{array}{cccccc}
{\rm FID} & {\rm Snum} & \alpha_{\rm x} & \log L &
{\rm FWHM_{Mg{\sc ii}}} & \sin i \\
(1) & (2) & (3) & (4) & (5) & (6) \vspace{5pt} \\
228 &   1   &   0.263  &   45.317  &   11289. &    0.788 \\
234 &   1   &   1.445  &   45.760  &   5957.9 &    0.495 \\
234 &  33   &   1.904  &   44.933  &   7597.9 &    0.427 \\
236 &   5   &   1.363  &   43.691  &   3058.9 &    0.066 \\
236 &  21   &   1.474  &   44.697  &   2799.2 &    0.114 \\
240 &  15   &   1.431  &   44.923  &   2054.1 &    0.085 \\
240 &  82   &   0.673  &   43.848  &   8635.6 &    0.282 \\
245 &   4   &   1.329  &   44.172  &   9330.3 &    0.363 \\
252 &   9   &   1.246  &   44.207  &   5762.9 &    0.213 \\
252 &  34   &   0.906  &   44.042  &   5500.9 &    0.184 \\
252 &  36   &   1.039  &   44.821  &   6023.7 &    0.311 \\
253 &   5   &   1.158  &   44.712  &   6279.0 &    0.308 \\
254 &  10   &   1.576  &   45.152  &   2672.3 &    0.141 \\
254 &  11   &   1.304  &   45.184  &   4100.3 &    0.242 \\
254 &  41   &   1.234  &   43.789  &   8023.5 &    0.251 \\
255 &  13   &   1.634  &   44.002  &   3943.3 &    0.117 \\
255 &  19   &   2.155  &   44.640  &   4988.5 &    0.227 \\
257 &  14   &   0.776  &   44.722  &   2289.8 &    0.087 \\
257 &  20   &   1.019  &   44.795  &   5742.9 &    0.291 \\
257 &  38   &   0.990  &   44.919  &   4387.0 &    0.227 \\
258 &   5   &   1.643  &   44.469  &   2210.5 &    0.069 \\
258 &  30   &   1.066  &   43.665  &   5031.0 &    0.132 \\
259 &   5   &   0.948  &   44.700  &   8445.0 &    0.426 \\
259 &   7   &   0.530  &   43.538  &   3213.5 &    0.063 \\
259 &  11   &   0.929  &   44.482  &   5777.5 &    0.248 \\
260 &   8   &   0.974  &   45.263  &   7374.4 &    0.487 \\
260 &  44   &   0.393  &   44.969  &   3458.0 &    0.176 \\
262 &   1   &   1.520  &   44.469  &   4756.4 &    0.196 \\
262 &  34   &   1.439  &   43.777  &   3921.5 &    0.101 \\
265 &  17   &   1.006  &   43.741  &   1942.8 &    0.028 \\
268 &  11   &   0.626  &   44.211  &   7908.2 &    0.308 \\
271 &   2   &   1.987  &   43.909  &   3078.8 &    0.078 \\
271 &   7   &   1.626  &   45.065  &   4282.0 &    0.239 \\
272 &   8   &   1.535  &   45.526  &   3480.2 &    0.241 \\
272 &  18   &   1.083  &   44.517  &   6079.5 &    0.268 \\
272 &  28   &   1.463  &   44.010  &   6463.0 &    0.219 \\
273 &   4   &   1.530  &   44.986  &   2648.9 &    0.126 \\
273 &  18   &   1.496  &   43.650  &   4464.5 &    0.111 \\
273 &  22   &   2.149  &   44.988  &   3691.0 &    0.192 \\
273 &  23   &   1.243  &   43.549  &   2658.8 &    0.046 \\
278 &   9   &   0.802  &   44.816  &   3814.6 &    0.182 \\
281 &  21   &   0.677  &   43.726  &   5329.8 &    0.147 \\
283 &   6   &   0.685  &   44.840  &   5194.1 &    0.265 \\
283 &  21   &   0.349  &   43.945  &   6278.3 &    0.204 \\
286 &   2   &   1.778  &   45.527  &   9763.7 &    0.748 \\
293 &   1   &   0.834  &   44.476  &   3177.5 &    0.118 \\
293 &  12   &   0.684  &   44.392  &   2488.9 &    0.079 \\
294 &   1   &   1.291  &   44.446  &   8196.2 &    0.362 \\
302 &  14   &   1.151  &   44.451  &   2202.9 &    0.068 \\
302 &  18   &   0.752  &   44.713  &   3272.7 &    0.141 \\
305 &  18   &   0.279  &   43.776  &   3081.2 &    0.071 \\
305 &  34   &   1.060  &   44.566  &   2122.8 &    0.070 \\
110 &  35   &   1.656  &   43.944  &   1884.0 &    0.033 \\
122 &   1   &   1.396  &   44.852  &   1083.0 &    0.024 \\
208 &  18   &   1.489  &   44.096  &   2639.2 &    0.070 \\
216 &  33   &   1.035  &   44.329  &   3176.6 &    0.107 \\
262 &  12   &   1.281  &   44.685  &   3120.1 &    0.131 \\
293 &  10   &   1.204  &   44.948  &   2972.2 &    0.143 \\
\end{array}
\]
\label{tab:rixosmgii}
\end{table}

\section{Relation between FWHM and $\alpha_{\rm X}$}
\label{sec:FWHM_ax}

\begin{figure}
	\psfig{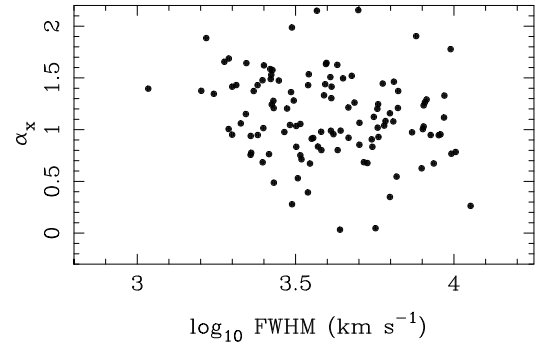}
	\caption{Observed $\alpha_{\rm X}$ vs FWHM$_{\rm Mg{\sc ii}}$. Note
	the weak anti-correlation between these two properties.}
	\label{fig:FWHM_ax}
\end{figure}
Several samples of AGN have shown an anti-correlation between FWHM and
$\alpha_{\rm x}$ \cite{BG92,Wandel98,P97} particularly for FWHM$_{\rm
H\beta}$. The viewing angle dependence of the FWHM leads us to expect
that $\alpha_{\rm X}$ might depend on $i$ also. Puchnarewicz et al.\
\shortcite{P97} also suggest that there is absorption of the soft
X-rays in the objects with broader lines, i.e. the more edge-on
objects. This might indicate a model in which the soft X-rays are
increasingly absorbed by some sort of torus of dust/gas as the
observer moves to a larger viewing angle. In the sample used here
there is in fact a weak correlation between $\alpha_{\rm X}$ and
FWHM$_{\rm Mg{\sc ii}}$ (fig.\ \ref{fig:FWHM_ax}).

\begin{figure*}
	\mbox{\psfig{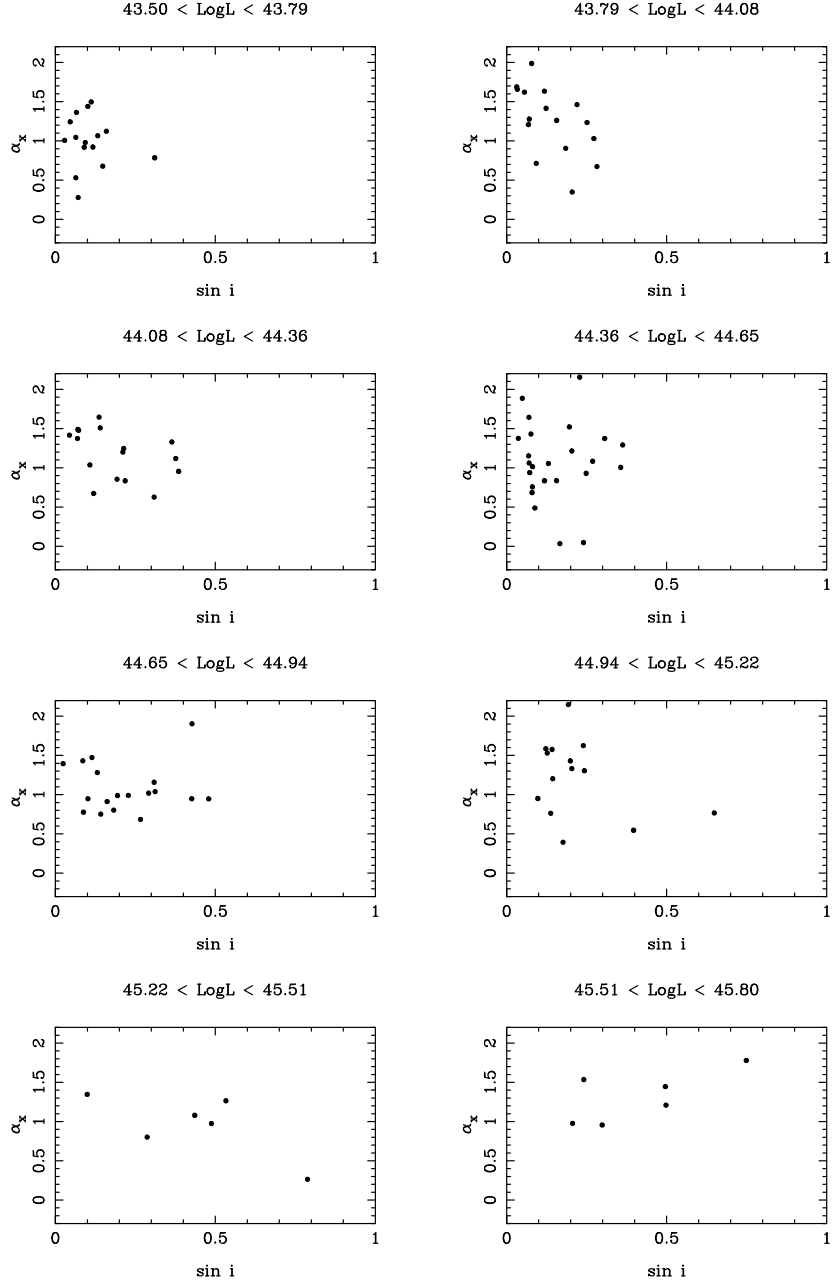}}
	\caption{Observed $\alpha_{\rm X}$ vs calculated $\sin i$
	for the objects with measured Mg{\sc ii} divided into
	luminosity bins. There is no obvious correlation, in either
	sense, common to all luminosity bins.}
	\label{fig:axsini}
\end{figure*}

We consider whether this correlation is at least in part due to the
relation between $i_{\ast}$ and $L$. Fig.\ \ref{fig:axsini} shows
$\alpha_{\rm X}$ plotted against $\sin i$ in luminosity bins. Notice
that their is no obvious anti-correlation between $\sin i$ and
$\alpha_{\rm X}$ with the data divided in this way. In fact some
luminosity bins indicate the opposite is true. This suggests that
viewing angle is not the primary parameter determining $\alpha_{\rm
X}$.

\section{Discussion}
\label{sec:discussion}

In this paper we have provided further evidence that the BLR in AGN is
axisymmetric. The case for axisymmetry in radio-loud systems has
always been strong with the observed correlations between line width,
$R$ and $\alpha_{\rm ox}$. With no such obvious measure of inclination
angle for radio-quiet systems the case is much harder to argue. The
RIXOS data used here is made up of both radio-loud and quiet
systems. There is no obvious failing of an axisymmetric model when
applied to all systems, as would be expected if the radio-quiets were
not axisymmetric. Also, with no clear difference between the
distribution of linewidths for radio-loud and radio-quiet systems, it
is reasonable to expect that the linewidths in radio-quiet systems
are, as in radio-loud systems, viewing angle dependent.

Marziani et al.\ \shortcite{Marziani96} argue the case for a
significant difference between the structure of the BLR in radio-loud
and radio-quiets systems. They show that the H$\beta$ line profiles
are predominantly redshifted and asymmetric in radio-louds whilst
being usually unshifted and symmetric in the radio-quiets. Conversely
C{\sc iv} is largely unshifted and symmetric in radio-louds and
blueshifted and asymmetric in radio-quiets. However, further
inspection of the distribution of these properties shows that they are
consistent with a single model for the BLR when we take into account
the effect of luminosity on opening angle. In the context of a
disc-wind model such as those of Cassidy \& Raine \shortcite{CR96} and
Chiang \& Murray \shortcite{Chiang96} we would expect the lines to be
blue shifted if the viewing angle is along (or close to) the
disc. This effect will be stronger in C{\sc iv} than in H$\beta$ which
tends to be produced close to the disc where the outward velocity
component is smaller and the outer clouds obscure the emission. This
obscuration may also cause us to see H$\beta$ predominantly from the
far side of the disc giving a tendency towards redshifted lines when
the opening angle is large. In general radio-loud sources are observed
at higher luminosities than radio-quiets and thus have a larger range
of possible viewing angles to the BLR. Thus we would expect to see a
blueshifted C{\sc iv} line in a smaller percentage of sources for
radio-louds than radio-quiets. The effect is increased by the fact
that in lower luminosity radio-quiets the opening angle is small and
the radial velocity component is significant at all viewing
angles. Thus, rather than showing that the radio-loud and radio-quiets
have different BLR structures, figure 4 of Marziani et al.\
\shortcite{Marziani96} can be interpreted as the change in the range
of those properties observed due to the effect of increasing opening
angle with luminosity.

We have shown that for the RIXOS sample \cite{P96,P97} our calculation
of the value of $\sin i$ for each system gives a realistic
distribution of angles. This distribution is, however, not uniform as
originally assumed \cite{RR98}. When only a small range of luminosity
is considered, it does become more uniform, but has a luminosity
dependent maximum value for $\sin i$. This result further supports the
assertion that the BLR is axisymmetric. In unified models it is
expected that the BLR, whether axisymmetric or not, can be viewed only
up to some maximum inclination, $i_{\ast}$, before being obscured. It
is also natural to expect that $i_{\ast}$ will be dependent upon
luminosity in the sense that higher luminosity systems will have
larger opening angles than low luminosity sources. In an axisymmetric
BLR model we have confirmed this result. Providing evidence for this
change in opening angle with luminosity has important implications for
unified models. Previously, the opening angle, at least for
radio-quiet AGN, has been estimated from the observed ratio of Seyfert
1s to Seyfert 2s. Our results show that this estimate needs to be
carried out at each luminosity rather than for complete samples.

We have also been able to show that the spectral index, $\alpha_{\rm
X}$, is not dependent primarily upon viewing angle. It is not
unreasonable to expect that the soft X-rays are obscured in more
edge-on systems giving a harder continuum. This was in fact predicted
from the RIXOS data by Puchnarewicz et al.\ \shortcite{P97}. Such a
viewing angle dependence would also explain the observed correlation
between FWHM$_{\rm H\beta}$ and spectral index observed in other
samples e.g. Boroson \& Green \shortcite{BG92}, where the broader
lines correspond to the harder X-ray spectrum. However, fig.\
\ref{fig:axsini} shows that when we plot $\alpha_{\rm X}$ against
$\sin i$ in luminosity bins we do not see the expected
anti-correlation between $\alpha_{\rm X}$ and $\sin i$. In fact at
some luminosities the data suggests that a positive correlation is
more likely. It appears that the observed anti-correlation is driven
at least partly by the dependence of $\sin i_{\ast}$ upon $L$, and is
not a consequence of an orientation dependent observed X-ray spectrum.

\section{Conclusions}

We have shown that if the kinematics of Mg{\sc ii} emission is
axisymmetric then the cone opening angle in the unified model is
dependent upon luminosity. The self-consistency of the picture
provides support for the view that the BLR is axisymmetric. As a
consequence we deduce that the X-ray spectral index is not primarily
dependent upon viewing angle.

\section{acknowledgements}

CMR acknowledges the support of PPARC, in the form of a research
studentship.


\begin{thebibliography}{}

	\bibitem[\protect\citename{Antonucci }1993]{Antonucci93}
	Antonucci R., 1993, ARA\&A, 31 ,473

	\bibitem[\protect\citename{Boroson }1992]{Boroson92} Boroson
	T.A., 1992, ApJ, 399, L15

	\bibitem[\protect\citename{Borson \& Green }1992]{BG92}
	Boroson T.A., Green R.F., 1992, ApJS, 80, 109

	\bibitem[\protect\citename{Boyle et al.\ }1988]{Boyle88}
	Boyle B.J., Shanks T., Peterson B.A., 1988, MNRAS, 235, 935

        \bibitem[\protect\citename{Boyle et al.\ }1993]{Boyle93}
        Boyle B.J., Griffiths R.E., Shanks T., Stewart G.C.,
        Georgantopoulos I., 1993, MNRAS, 260, 49

	\bibitem[\protect\citename{Boyle et al.\ }1994]{Boyle94}
	Boyle B.J., Shanks T., Georgantopoulos I., Stewart G.C.,
	Griffiths R.E., 1994, MNRAS, 271, 639

	\bibitem[\protect\citename{Brotherton }1996]{Brotherton96}
	Brotherton M.S., 1996, ApJS, 102, 1

	\bibitem[\protect\citename{Cassidy \& Raine }1996]{CR96}
	Cassidy I., Raine D.J., 1996, A\&A, 310, 49

        \bibitem[\protect\citename{Chiang and Murray}1996]{Chiang96}
        Chiang J., Murray N., 1996, ApJ, 466, 704

	\bibitem[\protect\citename{Corbin }1997]{Corbin97} Corbin
	M.R., 1997, ApJS, 113, 245

	\bibitem[\protect\citename{Dumont \& Colin--Souffrin
	}1990]{Dumont90} Dumont A-M., Collin--Souffrin S., 1990, A\&A, 
	229, 313

	\bibitem[\protect\citename{Gaskell }1982]{Gaskell82} Gaskell
	C.M., 1982, ApJ, 263, 79

	\bibitem[\protect\citename{Goad et al.\ }1999]{Goad99} Goad
	M.R., Koratkar A.P., Axon D.J., Korista K.T., O'Brien P.T.,
	1999, ApJL, 512, 95

	\bibitem[\protect\citename{Korista et al.\ }1995]{Korista95}
	Korista K.\ et al., 1995, ApJS, 97,285

	\bibitem[\protect\citename{Marziani et al.\ }1996]{Marziani96} 
	Marziani P., Sulentic J.W., Dultzin--Hacyan D., Calvani M.,
	Moles M., 1996, ApJS, 104, 37

	\bibitem[\protect\citename{Pei }1995]{Pei95} Pei Y.C., 1995
	ApJ, 438, 623

	\bibitem[\protect\citename{Puchnarewicz et al.\ }1996]{P96}
 	Puchnarewicz E.M., Mason K.O., Romero--Colmenero E.,
 	Carrera F.J., Hasinger G., McMahon R.G., Mittaz J.P.D., Page
 	M.J., Carballo R., 1996, MNRAS, 281, 1243

	\bibitem[\protect\citename{Puchnarewicz et al.\ }1997]{P97}
	Puchnarewicz E.M., Mason K.O., Carrera F.J. Brandt W.N.,
	Cabrera--Guera F., Carballo R., Hasinger G., McMahon R.,
	Mittaz J.P.D., Page M.J., Perez--Fouron I.,  Schwope A.,
	1997, MNRAS, 291, 177

	\bibitem[\protect\citename{Robinson }1995]{Robinson95}
	Robinson A., 1995, MNRAS, 272, 647

	\bibitem[\protect\citename{Rudge \& Raine }1998]{RR98} Rudge
	C.M., Raine D.J., 1998, MNRAS, 297, L1

	\bibitem[\protect\citename{Rudge \& Raine }1999]{RR99} Rudge
	C.M., Raine D.J., 1999, accepted for publication in MNRAS

	\bibitem[\protect\citename{Stirpe }1991]{Stirpe91} Stirpe
	G.M., 1991, Astron.\ Astrophys., 247, 3

	\bibitem[\protect\citename{Wandel \& Boller }1998]{Wandel98}
	Wandel A., Boller Th., 1998, A\&A, 331, 884

	\bibitem[\protect\citename{Wanders \& Horne }1994]{Wanders94}
	Wanders J., Horne K., A\&A, 1994, 289, 76

	\bibitem[\protect\citename{Wills \& Browne }1986]{Wills86}
	Wills B.J., Browne I.W.A., 1986, ApJ, 302, 56

	\bibitem[\protect\citename{Wills et al.\ }1993]{W93} Wills
	B.J., Brotherton M.S., Fang D., Steidel C.C., Sargent
	W.L.W., 1993, ApJ, 415, 563

	\bibitem[\protect\citename{Wills \& Brotherton
	}1995]{Wills95} Wills B.J., Brotherton M.S., 1995, ApJ, 448,
	L81

\end{thebibliography}
\end{document}